\documentstyle[12pt]{article}
\begin{document}
%------------------------------------------------------------
\begin{titlepage}
%------------------------------------------------------------
\title{ Gyroscopic Precession and Inertial Forces in Axially 
Symmetric Stationary Spacetimes }
%------------------------------------------------------------
\author{ K. Rajesh Nayak\thanks{ e-mail : nayak@iiap.ernet.in }
 \ and \ C.V. Vishveshwara\thanks{ e-mail : vishu@iiap.ernet.in }\\
Indian Institute of Astrophysics, Bangalore-560 034, India}
%------------------------------------------------------------
\date{}
%------------------------------------------------------------
\maketitle
%------------------------------------------------------------
\begin{abstract}

We study the phenomenon of gyroscopic precession and the 
analogues of inertial forces within the framework of general
relativity.  Covariant connections between the two are 
established for circular orbits in stationary spacetimes  
with axial symmetry. Specializing to static spacetimes,
we prove that gyroscopic precession and centrifugal force
both reverse at the photon orbits. Simultaneous non-reversal 
of these  in the case of stationary spacetimes is discussed.
Further insight is gained in the case of static spacetime by
considering the phenomena in a spacetime conformal to the original one. Gravi-electric and gravi-magnetic fields are 
studied  and their relation to inertial forces is 
established.
\end{abstract}   
%------------------------------------------------------------
\end{titlepage}
%------------------------------------------------------------
%------------------------------------------------------------
\section{Introduction}

Recently, two general relativistic phenomena apparently
related to each other in someway, have been investigated 
in considerable detail. These are gyroscopic precession 
and the general relativistic analogue of inertial forces.
Iyer and Vishveshwara \cite{IV} have given a 
comprehensive  treatment of gyroscopic precession in 
axially symmetric stationary spacetimes making use of the
elegant Frenet-Serret(FS) formalism. This forms the basis 
for a covariant  description of gyroscopic precession. At the
 same time, a general formalism defining inertial forces 
in general relativity has been presented by Abramowicz,
Nurowski and Wex \cite{ANW}.
	The motivation for this work stemmed 
from the earlier interest in centrifugal force and its
reversal. Such reversal in the Schwarzschild spacetime
at the circular  photon   orbit was first discussed by 
Abramowicz and Prasanna \cite{AP} and later in the case
of the Ernst spacetime by Prasanna\cite{P}. Abramowicz
\cite{A} showed that centrifugal force reversed at the
photon   orbit in all static spacetimes. He argued, on 
qualitative grounds, that gyroscopic  precession should
also reverse at the photon   orbit. Taking the Ernst 
spacetime as a specific example of static spacetimes Nayak
and Vishveshwara\cite{NV1} have shown that, in fact, both
centrifugal force and gyroscopic precession reverse at 
the photon   orbits. A similar study by Nayak and 
Vishveshwara \cite{NV2} in the Kerr-Newman spacetime 
indicates that the situation  in the case of stationary
 spacetimes is much more complicated than in the case of 
static spacetimes. Neither centrifugal force nor 
gyroscopic precession reverses at the photon   orbit.

	The above studies raise some interesting  
questions. Is gyroscopic precession directly related to
centrifugal force in all static spacetimes? If so, do they 
both necessarily reverse at  the photon   orbit? In the 
case of stationary spacetimes is it possible to make a
covariant connection between the gyroscopic precession
on the one hand and the inertial forces on the other, not
necessarily just the centrifugal force?  Does such a
connecting formula  reveal the individual
non-reversal of gyroscopic
precession and centrifugal force at the photon   orbit? In 
this paper we consider these and related questions.

	The contents of the present paper are organized in 
the following manner. In section 2 we sketch the 
Frenet-Serret description of gyroscopic precession. The 
formulae derived here form the basis for all subsequent 
considerations. They are further specialized to Killing 
trajectories in axisymmetric stationary spacetimes.
Although most of the contents of this section may 
essentially be found in previous papers ( \cite{IV}, 
\cite{NV1} and \cite{NV2}), they are included here for the
sake of completeness and ready readability of the present
paper. Furthermore, we present here formulae expressed in 
terms of Killing vectors which have not appeared before. 
Section 3 comprises the  formulation of general 
relativistic analogues of inertial forces and its 
specialization
to Killing trajectories. In section 4 covariant connections
are established between the precession and the   forces in 
both static and stationary spacetimes. Section 5 examines
the question of the reversal of gyroscopic precession in
relation to inertial forces in static and stationary metrics
in the light of these covariant connections.
In section 6 and 7 we consider two related aspects. First, we
treat gyroscopic precession as viewed  in a spacetime 
conformal to the original   static spacetime thereby factoring
out the contribution due to the gravitational force. Secondly,
we examine the idea of gravi-electric and gravi-magnetic 
fields in relation to the inertial forces. This is followed 
by concluding remarks of section 8.  
 
\section{Gyroscopic Precession}
\subsection{Frenet-Serret Description of Gyroscopic Precession}
The Frenet-Serret (FS) formalism offers a  covariant method of
treating gyroscopic precession. It turns out to be quite a
convenient and elegant description of the phenomenon when the 
worldlines along which the gyroscopes are transported follow 
spacetime symmetry directions or Killing vector fields.
In fact, in most cases of interest orbits corresponding to 
such worldlines are 
considered for simplicity. In the FS formalism the worldlines 
are characterized in an invariant geometric manner by defining along the curve  three parameters $\kappa$ the curvature and 
the two torsions $\tau_{1}$ and $\tau_{2}$ and an orthonormal tetrad. As we shall see,  the torsions $\tau_{1}$ and
 $\tau_{2}$ are directly related to gyroscopic precession.
All the above quantities can be expressed in terms of the 
Killing vectors and their derivatives. These considerations
 apply to a single trajectory in any specific example. 
However, additional geometric insight may be gained 
 by identifying the trajectory as a member of one or more 
congruences generated by combining different Killing vectors. For this 
purpose the FS formalism is generalized to what may be termed as 
quasi-Killing trajectories. For the sake of completeness we summarize 
below relevant formulae taken from reference    \cite{IV}.

	Let us consider a spacetime that admits a timelike Killing vector
$\xi^{a}$ and a set of spacelike Killing vector  $\eta_{(A)}$ 
(A=1,2,\ldots m). Then a quasi-Killing vector may be defined as 
\begin{equation}
\chi^{a} \equiv \xi^{a} + \omega_{(A)} \eta^{a}_{(A)}, 
\end{equation}
where (A) is summed over. The Lie derivative of the functions
$ \omega_{(A)} $ with respect to $ \chi^{a} $ is assumed to vanish,
\begin{equation}
{\cal L}_{\chi} \omega_{(A)} = 0.
\end{equation}
We adopt the convention that Latin indices $a,b,\ldots \ = \ 0-3 $ and
Greek indices $\alpha,\beta,\ldots \ = \ 1-3$ and the metric signature
is $(+,-,-,-)$. Geometrized units with $c=G=1$ are chosen. A congruence
of quasi-Killing trajectories is generated by the integral curves of
$\chi^a$. As a special case we obtain a Killing congruence when 
$\omega_{(A)}$ are constants.

	Assuming $\chi^a$ to be timelike, we may define the four 
velocity of a particle following $\chi^a$ by
\begin{equation}
e^{a}_{(0)} \ \equiv \ u^{a} \ \equiv \ e^{\psi} \chi^{a},
\end{equation}
so that
\begin{equation} 
e^{-2\psi}  = \chi^{a}\chi_{a}, \ \ \ \psi_{,a}\chi^{a} =0
\end{equation}
and 
\begin{equation}
\dot{e}_{(0)}^a \ \equiv \ e^{a}_{(0);b} e^{b}_{(0)} \ = \ F^{a}_{\ b } e^{b}_{(0)},
\end{equation}
where
\begin{equation}
F_{ab} \ \equiv \ e^{\psi} \left( \xi_{a;b} + \omega_{(A)} \eta_{(A)a;b}
\right).
\end{equation}
The derivative of $\omega_{(A)} $ drops out of the equation. The Killing
equation and the equation $ \xi_{a;b;c} \equiv R_{abcd}\xi^{d} $ 
satisfied by any Killing vector lead to
\begin{equation}
F_{ab} \ = \ -F_{ba} \  {\textstyle and }  \ \ \dot{F}_{ab} \ = \ 0 .
\end{equation}
 Now, the FS equations in general are given by
\begin{eqnarray}
\dot{e}^{a}_{(0)} & =  & \kappa e^{a}_{(1)},  \nonumber \\
\dot{e}^{a}_{(1)} & =  & \kappa e^{a}_{(0)} + \tau_{1} e^{a}_{(2)}, 
\nonumber \\
\dot{e}^{a}_{(2)} & =  & -\tau_{1} e^{a}_{(1)} + \tau_{2} e^{a}_{(3)}, 
\\
\dot{e}^{a}_{(3)} & =  & -\tau_{2} e^{a}_{(2)}, \nonumber
\end{eqnarray}
As mentioned earlier $\kappa$,$\tau_{1}$ and $\tau_{2}$ are respectively
the curvature, and the first and second torsions while $e^{a}_{(i)} $
form an orthonormal tetrad. The six quantities describe the worldline
completely. In the case of the quasi-Killing trajectories one can show 
that $\kappa$, $\tau_{1}$ and $\tau_{2}$ are constants and that each of 
$e^{a}_{(i)} $ satisfies a  Lorentz like equation:
\begin{equation}
\dot{\kappa} = \dot{\tau}_{1} = \dot{\tau}_{2} = 0,
\end{equation}
\begin{equation}
\dot{e}^{a}_{(i)} =  F^{a}_{ \ b}e^{b}_{(i)}.
\end{equation}
Further, $\kappa, \ \tau_{1}, \ \tau_{2} $ and $e^{a}_{(\alpha)}$
can be expressed in terms of $ e^{a}_{(0)} $ and  
$F^{n}_{\ ab} \equiv F_{a}^{\ a_{1}} F_{a_{1}}^{\ a_{2}} 
\cdots F_{a_{n-1}b}$.
\begin{eqnarray}
\kappa^{2}    &  =  &  F^{2}_{ab} e^{a}_{(0)} e^{b}_{(0)}  \nonumber \\
\tau^{2}_{1}  &  =  &  \kappa^{2} - \frac
{F^{4}_{ab} e^{a}_{(0)} e^{b}_{(0)}}{\kappa^{2}} \\
\tau^{2}_{2}  &  =  &   \frac {F^{6}_{ab} 
e^{a}_{(0)} e^{b}_{(0)}}{\kappa^{2}\tau^{2}_{1}} - \frac{ 
\left(\kappa^{2} - \tau^{2}_{1}\right)}{\tau^{2}_{1}}^{2} \nonumber
\end{eqnarray}
\begin{eqnarray}
e^{a}_{(1)}   &  =  &  \frac{1}{\kappa} F^{a}_{ \ b} e^{b}_{(0)} \nonumber \\
e^{a}_{(2)}   &  =  &  \frac{1}{\kappa \tau_{1}} \left[ F^{2a}_{ \ \ b}
- \kappa^{2} \delta^{a}_{b} \right] e^{b}_{(0)} \\
e^{a}_{(3)}   &  =  &  \frac{1}{\kappa \tau_{1} \tau_{2}} 
\left[F^{3a}_{\ \ b}  + \left( \tau^{2}_{1} - \kappa^{2}\right) 
F^{a}_{\  b}\right] e^{b}_{(0)} \nonumber
\end{eqnarray}
The above equations were first derived by Honig, Sch\"{u}cking and
Vishveshwara \cite{HSV} to describe charged particle motion in a homogeneous 
electromagnetic field. Interestingly, they are identical to
those that arise  in the case of quasi-Killing trajectories.

	Next let us consider an inertial frame of tetrad $\left( e^{a}_
{(0)}, f^{a}_{(\alpha)} \right) $ which undergoes Fermi-Walker (FW) 
transport along the worldline. The triad  $f_{(\alpha)}$ may be 
physically realized by a set of three mutually orthogonal gyroscopes. 
Then, the  angular velocity of the FS triad $ e^a_{(\alpha)} $ 
with respect to the FW triad $f^{a}_{(\alpha)}$ is
given by \cite{IV}
\begin{equation}
\omega_{FS}^a = \tau_{2} e_{(1)}^a +  \tau_{1} e_{(3)}^a.
\end{equation} 

	Or the gyroscopes precess with respect to the FS frame at a rate
given by $\Omega_{(g)} = -\omega_{FS} $.
In case of the Killing congruence $\omega_{FS} $ is identical
to the vorticity of the congruence.

Other important relations which relate the gyroscopic
precession and the acceleration and its derivatives can 
also be derived. The transport law for vector 
$ e^a_{(1)} \ \ ( \kappa e^a_{(1)}= a^a ) $ can be 
decomposed as 
\begin{equation}
\frac{de^a_{(1)}}{d\tau} = \left[ \left( a^a u^b - a^b
u^a \right) + \varepsilon^{cdab} u_c \omega_d \right]
e_{(1)b}
\end{equation}
Here $ ( a^au^b-a^bu^a ) $ gives the Fermi-Walker part.
The second term gives the spatial rotation  with
respect to the Fermi-Walker frame precessing with angular
velocity $\omega^a$  which is orthogonal to $u^a$,
\begin{equation}
u^a \omega_a = 0
\end{equation}
Since $\kappa e^a_{(1)} = a^a $ and $ \dot{\kappa} =0$,
We can show 
\begin{equation}
\dot{a}^a = \kappa^2 u^a + \varepsilon^{cdab} u_c \omega_d
a_b
\end{equation}
From this after some simplification we get 
\begin{equation}
\omega_p = \frac{1}{\kappa^2} \left[ \varepsilon_{pqra}
\dot{a}^a u^r a^q - a_p ( \omega_q a^q) \right]
\end{equation}
%------------------------------------------------------------
\subsection{Axially Symmetric Stationary Spacetimes}
%------------------------------------------------------------
	An axially symmetric stationary metric admits a timelike 
Killing vector $\xi^{a}$ and a spacelike Killing vector $\eta^{a}$ with
closed circular orbits around the axis of symmetry.  Assuming orthogonal
transitivity, in coordinates $\left( x^{0} \equiv t, \ \ x^{3} \equiv 
\phi \right) $ adapted to $\xi^a $ and $ \eta^a$ respectively the metric takes on its canonical form
\begin{equation}
ds^{2} \ = \ g_{00} dt^{2} \ + \ 2g_{03} dtd\phi \ + \ g_{33}d\phi^{2}
\ + \ g_{11} dr^{2} \ + \ g_{22}d\theta^{2} 
\end{equation}
with $g_{ab}$ functions of $ x^{1} \equiv r $ and $ x^{2} \equiv \theta
$ only. The quasi-Killing vector field 
\begin{equation}
\chi^{a} = \xi^{a} + \omega \eta^{a} 
\end{equation}
generates closed circular orbits around the symmetry axis with constant
angular speed $\omega$ along each orbit. The FS parameters and the 
tetrad can be determined either by the direct substitution of $\chi^{a}$
or by transforming to a rotating coordinate frame as discussed in \cite{IV}. They  can be written in 
terms of the Killing vectors and their derivatives as 
follows.
\begin{equation}
\kappa^{2}   \  =   \  -  g^{ab} a_a a_b
  \label{K1}
\end{equation}
\begin{equation}
\tau^{2}_{1}  \  =  \   
    \left[ g^{ab} a_a d_b \right]^2 \label{T1}
\end{equation}
\begin{equation}  
\tau^{2}_{2}  \  =   
\left[ \frac{\varepsilon^{abcd}}{\sqrt{-g}}
 n_a \tau_b a_c  d_d
\right]^2 \label{T2}
\end{equation}
\begin{eqnarray}
e^{a}_{(0)}    &   =   &   \frac{1}{ \sqrt{{\cal A}}} ( 1,0,0,\omega)
\nonumber \\
e^{a}_{(1)}    &   =   &   -\frac{1}{\kappa} (0 , g^{11} a_1 , g^{22} a_2 ,0) \nonumber \\
e^{a}_{(2)}    &   =   &   \frac{1}{ \sqrt{{\cal A}} \sqrt{ -\Delta_{3}}}
( {\cal B} , 0,0,-{\cal C}) \\
e^{a}_{(3)}    &   =   &    \frac{ \sqrt{ g^{11} g^{22}} }{  \kappa }
(0,-a_2,a_{1},0) \nonumber
\end{eqnarray}
In the above,
\begin{eqnarray}
 d_a & = & \left( \frac{ {\cal B}}{2 \sqrt{-\Delta_3} \kappa}
\right) 
 \left[ \frac{ {\cal B}_a}
{{\cal B}} - \frac{ { \cal A}_a}{{\cal A}} \right]  
\ \ = \ \ \left( \frac{ {\cal B}}{\sqrt{-\Delta_3} \kappa}
\right) \ [ b_a - a_a ] \nonumber
\\ 
a_a & = & \frac{ {\cal A}_a}{2 {\cal A}} \nonumber \\
b_a & = & \frac{ {\cal B}_a}{2 {\cal B}} \nonumber \\
{\cal A}       &  =  &  (\xi^a \xi_a) \  + \ 2\omega \ (             \eta^a 
\xi_a ) \ + \ \omega^{2} \ (\eta^a \eta_a) \nonumber \\ 
 {\cal B}       &  =  &  (\eta^a \xi_a) \ + \ \omega \ ( \eta^a \eta_a )
 \label{A:B} \\
{\cal C}       &  =  & (\xi^a \xi_a) \  + \ \omega \ (\eta^a \xi_a )
 \nonumber \\
{\cal A}_{a} &  =  & (\xi^b \xi_b) _{,a} + 2 \omega 
(\eta^b \xi_b)_{,a} +\omega^{2} (\eta^b \eta_b)
_{,a} ; \ \ \ a = 1,2.  
 \nonumber \\
{\cal B}_{b} &  =  & (\eta^a \xi_a)_{,b} +
\omega (\eta^a \eta_a)_{,b} ; \ \ \ b=1,2.  \nonumber \\       \Delta_{3}     &  =  & (\xi^a \xi_a)(\eta^b \eta_b) \ -  \ 
(\eta^a \xi_a)^2 \nonumber 
\end{eqnarray}
where  $n^a $ is the unit vector along 
$\zeta_a = \xi_a - \frac{\left( \xi^b\eta_b \right)}{\left( \eta^c\eta_c
\right) } \eta_a $ and  $ \tau^i $ is the unit vector along 
the rotational Killing vector $\eta^a$.
We may note that all the above equations can be specialized 
to a static spacetime by setting $ \xi^a \eta_a = 0 $ 
and $\zeta^a \equiv \xi^a $. 
%------------------------------------------------------------
%------------------------------------------------------------
\section{Inertial Forces}
%------------------------------------------------------------
%------------------------------------------------------------
%------------------------------------------------------------
\subsection{General Formalism}
	As has been mentioned earlier, in a recent paper Abramowicz 
{\it et. al.} \cite{ANW} have formulated the general relativistic  analogues  of
inertial forces in an arbitrary spacetime.  The particle
four velocity $u^a$ is decomposed as
\begin{equation}
u^a = \gamma \left( n^a + v \tau^a \right) \label{DEC}
\end{equation}
In the above, $n^a$ is a globally hypersurface orthogonal timelike unit
vector, $\tau^a$ is the unit vector orthogonal to it along which the 
spatial three velocity  $v$ of the particle is aligned and $\gamma$ is the 
normalization factor that makes $u^au_a=1$. 

	Then the forces acting on the particle are written down as
\begin{eqnarray}
{\textstyle Gravitational \ \ \ force} \ \  G_{k} & = & \phi,_k \nonumber \\
{\textstyle Centrifugal \ \  \ force } \ \ Z_{k} & = & - \left( \gamma v \right)^2
\tilde{\tau}^i \tilde{\nabla}_i \tilde{\tau}_k \nonumber \\
{\textstyle Euler  \ \ \ force } \ \  E_k & = &  -\dot{V} \tilde{\tau}_k
\nonumber \\
{\textstyle Coriolis-Lense-Thirring  \ \ \ force } \ \ C_k & = & \gamma^2 v X_k
\end{eqnarray}
where 
\begin{eqnarray}
\dot{V} & = & (v e^{\phi} \gamma )_{,i} u^i \nonumber \\
X_k & = & n^i \left( \tau_{k;i} - \tau_{i;k} \right)  \\
\phi_{,k} & = &  -n^i n_{k,i}  \nonumber
\end{eqnarray}
Here $\tilde{\tau}^i $ is the unit vector along $\tau^i$ in the 
conformal space orthogonal to $n^i$ with the metric 
\begin{equation}
\tilde{h}_{ik} = e^{-2\phi} \left( g_{ik} - n_in_k \right)
\end{equation}
One can show that the covariant derivatives in the two spaces are 
related by 
\begin{equation}
\tilde{\tau}^{i} \tilde{\nabla}_{i} \tilde{\tau}_{k} =
 \tau^{i} \nabla_{i}  \tau_{k} - \tau^{i} \tau_{k} \nabla
_{i} \phi - \nabla_{k} \phi.
\end{equation}
We shall now apply this formalism to axially symmetric 
stationary space-times.
%-----------------------------------------------------------
\subsection{Inertial Forces in Axially Symmetric Stationary 
Spacetimes}
%-----------------------------------------------------------
%
As has been shown by Greene, Sch\"{u}cking and Vishveshwara
\cite{GSV}, axially symmetric stationary spacetimes with
orthogonal transitivity admit a globally hypersurface    
orthogonal timelike vector field 
\begin{equation}
\zeta^a  = \xi^a + \omega_0 \eta^a, 
\end{equation}
where the fundamental angular speed of the irrotational 
congruence is
\begin{equation} 
\omega_0 \ = \ -(\xi^a \eta_a ) / (\eta^b
\eta_b) \label{OM}
\end{equation}
  The unit vector along $\zeta ^a$ is identified 
with $n^a$. Further, if  $u^a$ follows a quasi-Killing 
circular trajectory, then $\tau^i$ is along the rotational 
Killing vector  $\eta^a$. In this case it is easy to show that
$\dot{V}=0$  and hence the Euler force does not exist.

More specifically, 
\begin{eqnarray} 
u^a \ = \  e^{\psi} ( \xi^a + \omega \eta^a ) 
\ = \ e^{\psi} \chi^a \ 
\  = \  \gamma ( n^a + v \tau^a ). 
\end{eqnarray}
Then we have
\begin{eqnarray}
n^a & = &  e^{-\phi } \zeta^a \\ \nonumber
\tau^a & = & e^{-\alpha} \eta^a \\ \nonumber
\gamma & = & e^{\psi + \phi }  \label{NA} 
\\
v & = & e^{ -\phi + \alpha } \left( \omega -  \omega_0
 \right) \nonumber
\end{eqnarray}
where 
\begin{equation}
\phi \ = \  \frac{1}{2} \ln(\zeta^a \zeta_a), \ \ \ 
\alpha \ = \ \frac{1}{2} \ln(-\eta^a\eta_a), \ \ \ 
\psi \ = \ \frac{1}{2} \ln(\chi^a\chi_a) \ \ \  
\end{equation} 
From the above relations, we can write down the inertial 
forces from their definitions as follows. \newline
Gravitational force    
\begin{equation}
G_k \ = \ \phi_{,k}
\end{equation}
Centrifugal force 
\begin{equation}
Z_k  =  \frac{1}{2} e^{2(\psi+\phi)} \tilde{\omega}^2
\left( \frac{  \eta^a \eta_a }{ \zeta^b \zeta_b} \right)
_{,k} 
\label{ZZZ} 
\end{equation}
Coriolis-Lense-Thirring  force 
\begin{equation}
C_k  =  e^{2(\psi+\alpha)} \tilde{\omega}
\left( \frac{ \xi^a \eta_a}{\eta^b\eta_b} \right) _{,k}  
\end{equation}
where $ \tilde{\omega} = \left( \omega - \omega_0 \right)
$.
 For a particle following a quasi-Killing trajectory, 
 inertial forces  are  
proportional to   gradiants of  functions.
 %---------------------------------------------------------
\subsection{Specialization to Static Spacetimes:}
%----------------------------------------------------------
In a static spacetime the global timelike Killing vector 
$\xi^a$  itself is hypersurface orthogonal. The unit vector
$n^a$ is now aligned along $\xi^a$,
\begin{equation}
n^a \ = \ e^{-\phi} \xi^a. 
\end{equation}
Then we have the inertial forces as follows: \newline
Gravitational force
\begin{equation}
G_k \ = \  \phi_{,k}
\end{equation}
where  $ \phi \ = \ \frac{1}{2}  ln(\xi^a \xi_a ) $ \newline
Centrifugal force

\begin{equation}
Z_k \ = \  - \frac{\omega^2}{2} e^{2(\psi+\alpha)} \
\left[\ln\left( \frac{\eta^i\eta_i} {\xi^j\xi_j} \right) \right]_{,k}
\label{ZS}
\end{equation}
Coriolis-Lense-Thirring force is identically zero,
\begin{equation}
C_k = 0 
\end{equation}
%-----------------------------------------------------------
%------------------------------------------------------------
\section{Covariant Connections}
%------------------------------------------------------------
%------------------------------------------------------------
In the preceding section  we have  derived  
expressions for $\tau_1$  and $\tau_2$ which give gyroscopic
precession rate in terms of the Killing vectors.
  Similarly, inertial forces in an arbitrary 
axisymmetric stationary spacetime have also been written down
in terms of the Killing vectors. All these quantities have 
been defined in a completely covariant manner. We shall now
proceed to establish  covariant connections between 
gyroscopic precession, {\it i.e.} the FS torsions $\tau_1 $
and $ \tau_2 $, on the one hand and the inertial forces on
the other. First, we shall consider the simpler case 
of static       spacetimes.
%-----------------------------------------------------------
\subsection{Static Spacetimes}
%-----------------------------------------------------------
  We have derived in equation 
(\ref{T1}) and
(\ref{T2}), the FS torsions  $\tau_1$ and $\tau_2$  
for a stationary spacetime.
 As has been mentioned earlier, for a static spacetime
$\xi^a \eta_a=0$  and $\zeta^a = \xi^a$  in the above 
equations as well as in the expressions for inertial  
forces.
 With this specialization, centrifugal
force can be written from equation (\ref{ZS}) as 
\begin{equation}
Z_b   \ = \ e^{-(\phi-\alpha)} \omega \kappa  d_ b 
\label{EQN47}
\end{equation}
Substituting equation (\ref{EQN47}) in equations (\ref{T1}) 
and (\ref{T2})
we arrive   at the relations
\begin{equation}
\tau_1^2 \ = \ \frac{\beta^2}{\omega^2} 
\left[ a^b Z_b \right]^2 \label{conection}
\end{equation}
and 
\begin{equation}
\tau^2_2 \ = \ \frac{\beta^2}{\omega^2}
\left[ \frac{\varepsilon^{abcd}}{\sqrt{-g}} n_a \tau_b a_c Z_d \right]^2
\label{conection1}
\end{equation}
where 
\begin{equation}
\beta \ = \ \frac{e^{(\phi-\alpha)}}{\kappa}
\end{equation}
The equations above relate gyroscopic precession directly 
to the centrifugal       force. The two torsions $\tau_1$
and $\tau_2$, equivalent to the two components of precession,
are respectively proportional to the scalar and cross 
products of acceleration and the centrifugal force.
 We shall discuss the consequences
of   these relations later on.
%----------------------------------------------------------
\subsection{Stationary       Spacetimes}     
%-----------------------------------------------------------
From equation (\ref{A:B}) we have 
\begin{eqnarray}
{\cal A}       &  =  &  (\xi^a \xi_a) \  + \ 2\omega \ (             \eta^a 
\xi_a ) \ + \ \omega^{2} \ (\eta^a \eta_a) \nonumber \\
{\cal B}       &  =  &  (\eta^a \xi_a) \ + \ \omega \ ( \eta^a \eta_a ) \nonumber 
\end{eqnarray}
We decompose the angular speed $\omega $ with reference   to
the fundamental angular speed of the irrotational 
congruence $ \omega_0 = -\frac{(\xi^a \eta_a)}{(\eta^a 
\eta_a )} $,
\begin{equation}
\omega = \tilde{\omega} + \omega_0.
\end{equation}
Then we have 
\begin{eqnarray}
{\cal A} & = & \zeta^a \zeta_a \ + \ \tilde{\omega}^2 \ 
\eta^a \eta_a \nonumber \\
{\cal B} & = & \tilde{\omega} \ \eta^a \eta_a 
\label{A:B1}
\end{eqnarray}
Similarly, we get 
\begin{eqnarray}
{\cal A}_a & = & (\zeta^b \zeta_b)_{,a} \ + \ 2 
\tilde{\omega} \  {\cal C}_a \ + \ \tilde{\omega}^2 \
(\eta^b \eta_b )_{,a} \nonumber \\
{\cal B}_a & = & {\cal C}_a \ + \ \tilde{\omega} \
(\eta^b \eta_b )_{,a}
\label{A:B2} 
\end{eqnarray}
where
\begin{equation}
{\cal C}_a \ \equiv \ (\xi^b \eta_b)_{,a} \ + \ \omega_0 \ 
 (\eta^b \eta_b )_{,a} 
\end{equation}
or equivalatly
\begin{equation}
{\cal C}_a \ =  \  - ( \xi^b \eta_b ) \omega_{0,a}
\end{equation}
From equation (\ref{A:B}),(\ref{A:B1}) and (\ref{A:B2})
we can show 
\begin{equation}
d_a \ = \ -e^{2\psi}\frac{e^{-(\phi+\alpha)}\tilde{\omega} }{2 \kappa} 
\left\{
(\zeta^p\zeta_p)  {\cal C}_a \ + \ \tilde{\omega} 
\left[ (\zeta^p\zeta_p) (\eta^q \eta_q)_{,a} \ - \
(\eta^p\eta_p)(\zeta^q\zeta_q)_{,a} \right] \ - \
\tilde{\omega}^2 \ (\eta^p \eta_p)  {\cal C}_a
\label{D1}
\right\}
\end{equation}
Further, it is is easy to see that ${\cal C}_a $ is directly
proportional to the Coriolis force,
\begin{equation}
{\cal C}_a \ = -\ e^{-2\psi } \ \tilde{\omega}^{-1} C_a
\label{CCC}
\end{equation}
where $C_a$ is the Coriolis-Lense-Thirring force. Then   equation (\ref{D1}) takes on the form
\begin{equation}
 d_a \ = \
\frac{ e^{(\phi-\alpha)}}{\tilde{\omega} \kappa }
\left\{   Z_a \ - \  \frac{1}{2} \left[ 1 +
\tilde{\omega}^2  e^{2(\alpha-\phi)}  \right]  C_a \right\}
\end{equation}
Where $Z_a$ is the centrifugal force.

Substituting this in equation(\ref{T1}) for $\tau_1^2$ 
we get the relation,
\begin{equation}
\tau^2_1 \ = \ \frac{\beta^2}{\tilde{\omega}^2} 
\left[ g^{ab} a_a \left( Z_b \ + \ \beta_1 C_a \right)
\right]^2 \label{ZZZ1}
\end{equation}
where
\begin{eqnarray}
\beta & = & \frac{e^{( \phi-\alpha )}}{\kappa}
\nonumber \\
\beta_1 & = & -\frac{1}{2} \left[ 1 +  \tilde{\omega}^2  e^{2(\alpha-\phi)}  \right] 
\end{eqnarray}
Again, from equation (\ref{T2}) and (\ref{D1}), we obtain
the expression 
\begin{equation}
\tau^2_2 \ = \ \frac{\beta^2}{\tilde{\omega}^2} \left[
\frac{\varepsilon^{abcd}}{\sqrt{-g}} n_a \tau_b a_c  \left( Z_d  \ + \ \beta_1
C_d \right) \right]^2 \label{ZZZ2}
\end{equation} 
These relations are more complicated than those we have 
derived in the static case. Nevertheless, they closely 
resemble the latter with the centrifugal force replaced by
the combination of centrifugal and Coriolis forces
$ ( Z_a + \beta_1 C_a ) $. 
The static case formulae are obtained from those of 
stationary case 
  by setting the Coriolis force to zero.

A formula for gyroscopic precession in the case of circular
orbits in axially symmetric stationary spacetimes was 
derived  by Abramowicz, Nurowski and Wex\cite{ANW2}  within a
different framework. We note that gyroscopics precession does
not involve the gravitational force. In case of geodetic  orbits,
total force is zero but not the centrifugal and Coriolis force
individually. Therefore gyroscopic precession is also nonzero
even for geodetic orbits.

	We may also note that the definition of inertial force
is not unique. For instance,  De Felice\cite{DEF1}, Semerak\cite{OS1}
 Barrabes, Boisseau and Israel \cite{BBI} have  suggested approaches
different from the one employed in the present paper. The gyroscopics
precession was solved in an alternative formalism by Semerak\cite{OS2,OS3}
(see also De Felice\cite{DEF2} ). A recent paper by Bini, Carini and
Jantzen\cite{BCJ1}  has established  the geometrical basis and 
interrelation of  various modes of force splitting while they apply their
general results to circular orbits in axially symmetric stationary
spactimes\cite{BCJ2}.

%----------------------------------------------------------
%----------------------------------------------------------
\section{Reversal of Gyroscopic Precession and Inertial 
Forces:}
%----------------------------------------------------------
%----------------------------------------------------------
The condition for the reversal of gyroscopic precession
is given by
\begin{equation}
\omega^a_{FS} \ = \ \tau_1 e^a_{(3)} \ + \ \tau_2 e^a_{(1)} \
= \ 0
\end{equation}
Since $e^a_{(1)}$ and $e^a_{(3)}$  are linearly independent 
vector fields at each point, this condition is the same 
as requiring 
\begin{equation}
\tau_1 \ = \ \tau_2 \ = \ 0 
\end{equation}
By considering the actual structure of $ \tau_2 $,
it is easy to show that $ \tau_2 $ becomes zero on a plane 
about which the metric components are reflection invariant.
The equatorial plane in the black hole spacetime is an example of this.

	We shall now examine the vanishing of the FS 
torsions  expressed as above and in relation to the inertial forces.
%----------------------------------------------------------
\subsection{Static Spacetimes}
%----------------------------------------------------------
In what follows, we shall prove a theorem that relates the 
simultaneous reversal of centrifugal force and gyroscopic
precession to the existence of a null circular orbit.
	We start from the condition for gyroscopic precession 
reversal, {\it i.e.} $ \tau_1 \ = \ \tau_2 \ =
\ 0 $, and show that  at the point where this occurs a null 
circular geodesic must exist.

Setting  $\tau_2=0$  in equation (\ref{T2}) and  noting that the only
 nonvanishing components of $n_a$ and $\tau_a$ are respectively
$n_0$ and $\tau_3$ we arrive at the condition  
\begin{equation}
{\cal A}_1 {\cal B}_2 \ = \ {\cal A}_2 {\cal B}_1 
\label{T2=0}
\end{equation}
Further setting $ \tau_1 =0  $ in 
equation (\ref{T1}), we obtain     
\begin{equation}
g^{11} \left( \frac{ {\cal A}_1 {\cal B}_1 }{\cal B} -
\frac{{\cal A}^2_1}{\cal A} \right) \ + \ 
g^{22} \left( \frac{ {\cal A}_2 {\cal B}_2 }{\cal B} -
\frac{{\cal A}^2_2}{\cal A} \right) \ = 0.
\label{T1=0}
\end{equation}
We shall now assume that the gyroscope is transported  along
a circular orbit which is not a geodesic, {\it i.e.} 
$\kappa \neq 0$. This we do in anticipation of the result
that a null geodesic, not a timelike one, exists with its
spatial trajectory identical to that of this timelike orbit.
Now $\kappa \neq 0 $ implies ${\cal A}_1 \neq 0 $ and 
${\cal A}_2 \neq 0 $  from equation  (\ref{A:B}). Then from
equation (\ref{T2=0}) and (\ref{T1=0}) we arrive at
\begin{equation}
{\cal AB}_1 - {\cal BA}_1 \ = \ 0 
\end{equation}
and 
\begin{equation}
{\cal AB}_2 - {\cal BA}_2 \ = \ 0 
\end{equation}
Combining the above two  equations,
\[
{\cal AB}_a - {\cal BA}_a \  = \  0  
\]
Then, equation (\ref{A:B}) reduces this condition to 
\begin{equation}
(\xi^b\xi_b)(\eta^c\eta_c)_{,a} \ - \ (\xi^b\xi_b)_{,a}
(\eta^c\eta_c) \ = \   0 
\label{PRC}
\end{equation}
With the help of this equation we can show that, if a
circular geodesic exists where precession reverses, then
it has to be null as follows.

The condition for circular geodesic is
\begin{equation}
(\xi^b \xi_b )_{,a} + \omega^2 (\eta^b\eta_b)_{,a} \ = \ 0
\end{equation}
This can be proved from the geodesic equation, assuming that
the four velocity $u^a$ is proportional to $ \xi^a + \omega
\eta^a$.
Using condition (\ref{PRC}), this reduces to
\begin{equation}
\frac{(\xi^b\xi_b)_{,a}}{(\xi^c\xi_c)} \left[
(\xi^d\xi_d) \ + \ \omega^2 (\eta^d\eta_d) \right] = 0  
\end{equation}
Since $\frac{(\xi^b\xi_b)_{,a}}{2(\xi^c\xi_c)}$   
 is the  gravitational force, which is assumed to be nonzero,  
this is equivalent to
\begin{equation}
(\xi^d\xi_d) \ + \ \omega^2 ( \eta^d\eta_d) \ = \ 0
\end{equation}
This means that   the geodesic, if one exists, is null. Now we shall
show  that in fact  a 
geodesic must exist at the point of precession reversal.

If a geodesic does not exist at the point of reversal, then
\begin{equation}
(\xi^b\xi_b)_{,a} \ + \ \omega^2 (\eta^b \eta_b)_{,a} 
\ \neq   \ 0
\end{equation}
for all values of $\omega$.  However, equation (\ref{PRC})
may be recast as
\begin{equation} 
(\xi^b\xi_b)_{,a} \ - \left(\frac{\xi^c\xi_c}{\eta^c\eta_c}
\right)  (\eta^b \eta_b)_{,a}  \ = \ 0.
\end{equation}
This shows that the geodesic condition is satisfied for 
$\omega^2 \ = \  - \left(\frac{\xi^c\xi_c}{\eta^c\eta_c}
\right) $. Therefore there does exist a geodesic and we have 
already shown that it has to be null.
	We shall now prove the converse, {\it i.e.} if a
circular null geodesic exists then $\tau_1$ and $\tau_2 $
are zero at the null geodesic. 

The condition for a circular null geodesic is given by 
equation (\ref{PRC}). 
Dividing this equation by $(\xi^a \xi_a)(\eta^b \eta_b)$, we 
see that it reduces to  $\left[\ln\left(\frac{\eta^b \eta_b}
{\xi^c \xi_c} \right) \right]_{,k}$ which is proportional
to $Z_a$ from equation (\ref{ZS}) and is equal to zero.
 Further, from the dependence
of $\tau_1$ and $\tau_2 $ on  $Z_a$ from equations (\ref{conection})
and (\ref{conection1})  we see that 
  $ \tau_1 \ = \ \tau_2 \ = \ 0 $.
 We may note the fact that both gyroscopic precession and 
centrifugal force reverse simultaneously  as is evident from
equations (\ref{conection})
and (\ref{conection1}). 
We have therefore proved the following theorem. 
\vspace{0.5cm}
\newline
{\it Theorem: } In the case of 
 circular orbits in static spacetimes
reversal of gyroscopic precession and centrifugal force
 takes place at some 
point, if and only if a null geodesic exists at that point.
\vspace{0.5cm}
\newline
%----------------------------------------------------------
\subsection{Stationary Spacetimes}
%----------------------------------------------------------
In section 4 we have derived expressions for $\tau_1$
and $\tau_2$, that embody gyroscopic precession, in terms
of inertial forces, namely the centrifugal force $Z_a$
and Coriolis-Lense-Thirring force $C_a$. These are
complicated expressions and $\omega$ does not stand out
as an overall multiplicative coefficient. Consequently,
 reversal of gyroscopic precession is not related directly
 to that of these forces individually.  As has been discussed
in reference   \cite{NV2}, these reversals occur at different
places and also not at the null geodesic. 
Nevertheless, one can
see from equation (\ref{ZZZ1}) and (\ref{ZZZ2}) that 
gyroscopic precession reverses at a point where the 
combination of centrifugal and Coriolis forces given by
$ (Z_a + \beta_1 C_a ) $ goes to zero.

 We shall
derive the angular velocity of  a timelike orbit whose three
dimensional  trajectory coincides with a null geodesic in
terms of inertial   forces. Although there are no reversals
at the null geodesic, this should give an idea of how
these forces are structured along the null trajectory.

Conditions for the existence of circular null geodesic are 
\begin{equation}
{\cal A}       \  \equiv  \  (\xi^a \xi_a) \  + \ 2\bar{\omega} \ (             \eta^a 
\xi_a ) \ + \ \bar{\omega}^{2} \ (\eta^a \eta_a) \ = \ 0
\end{equation}
and
\begin{equation}
{\cal A}_{a} \  \equiv  \ (\xi^b \xi_b) _{,a} + 2 \bar{\omega} 
(\eta^b \xi_b)_{,a} +\bar{\omega}^{2} (\eta^b \eta_b)
_{,a} \ = \ 0  \label{AaNULL}
\end{equation}
The expression for ${\cal A}$ can also be written as
\begin{equation}
{\cal A} \ = \ \zeta^a\zeta_a \ + \ \tilde{\bar{\omega}}^2 
\eta^a \eta_a,
\end{equation}
where
\begin{equation}
\bar{\omega} \ = \ \tilde{\bar{\omega}} - \frac{ \xi^b\eta_b}
{\eta^c \eta_c}
\end{equation}
Then ${\cal A} \ = \ 0 $ implies
\begin{equation}
\tilde{\bar{\omega}} \ = \ \pm \sqrt{ - \frac{\zeta^a \zeta_a}{
\eta^a \eta_a} }
\end{equation}
Further, from equation (\ref{AaNULL})
\begin{equation}
{\cal A}_a \ = \ \frac{1}{\eta^p \eta_p} \left\{
(\eta^q \eta_q) (\zeta^b\zeta_b)_{,a} \ -  \  
(\zeta^b\zeta_b)(\eta^q \eta_q)_{,a} \right\} \  \pm \
\sqrt{ - \frac{\zeta^b \zeta_b}{
\eta^q \eta_q} } {\cal C}_a \ = \ 0
\label{NUL} 
\end{equation}
This has to be zero for a null geodesic. For a timelike
curve with the same spatial orbit, but having 
 angular velocity
$\tilde{\omega}$ with respect to $n^a$,we have
 from equation 
(\ref{ZZZ}) and (\ref{CCC}) 
\begin{equation}
 Z_k \ = \  \frac{ e^{2\psi}}{2} \ \left( \omega - \omega_0 \right)^2
\ \frac{1}{ \zeta^a \zeta_a } \
\left[ \ (\zeta^b \zeta_b) \ (\eta^c \eta_c)_{,k} \ - \
(\eta^b \eta_b ) \ (\zeta^c \zeta_c)_{,k} \right]  
\end{equation}
\begin{equation}
{\cal C}_a = -e^{-2\psi} \tilde{\omega}^{-1} C_a 
\end{equation}
Substituting in equation (\ref{NUL}), we get
\begin{equation} 
\frac{1}{\eta^p \eta_p} \left\{ 2 e^{-2\psi}
  (\zeta^p\zeta_p) \tilde{\omega}^{-2} Z_a \right\} \ \mp  \ \left\{ 2\sqrt{ - \frac{\zeta^b \zeta_b}{
\eta^q \eta_q} }  e^{-2\psi}\tilde{\omega}^{-1} C_a 
\right\} \ =  \ 0 
\end{equation}
This  reducess to the equation
\begin{equation}
 \sqrt{ - \frac{\zeta^p \zeta_p}{
\eta^q \eta_q} } Z_a \ \mp  \tilde{\omega} C_a \ =  \ 0 
\end{equation}
which gives  $ \tilde{\omega} $ in terms of centrifugal and
Coriolis forces.
%----------------------------------------------------------
%----------------------------------------------------------
\section{Gyroscopic Precession and Inertial Forces in 
Conformal Static Spacetimes}
%----------------------------------------------------------
%----------------------------------------------------------
Some further insight into gyroscopic precession and inertial
forces may be gained by considering them in a space 
conformal to the original one as given in Abramowicz, Carter 
and Lasota \cite{ACL}.
 In the case of the static
metric we carry out the conformal transformation 
\begin{equation}
\hat{g}_{ab} \ = \ e^{-2\phi} g_{ab}
\end{equation}
If we choose
\begin{equation}
  e^{-2\phi} \ = \ g^{00} \ = \ \frac{1}{g_{00}} \ = \ 
\frac{1}{\hat{\phi} } \label{CON}
\end{equation}
then, $ \hat{g}_{00} = \hat{g}^{00} = 1 $.
Spatial part of metric  $ \hat{g}_{ab}$  corresponds to
optical geometry defined in reference
\cite{ACL} for identifying inertial forces in such geometry.
 Purely in the 
conformal space, with out referring to original $ g_{ab}$,
we have
\begin{equation}
\hat{u}^a \hat{\nabla}_a \hat{u}^b \ = \ 0 \label{GEO}
\end{equation}
for a stationary observer with four velocity $ \hat{u}^a =
(1,0,0,0) $, where $ \hat{\nabla}_a $ is covariant
derivative with respect  to the conformal metric  $ \hat{g_{ab}}$.
The two four velocities $u^a $ and $\hat{u}^a $ are related by
$u^a = e^{-\phi} \hat{u}^a $.
Equation (\ref{GEO}) indicates that because of dilation,
 $ \hat{u}^a $ follows a geodesic trajectory in the conformal
metric. This is equivalent to the statement that the only force 
acting on a particle at rest in the original space is 
gravitational force which is not felt in the  conformal   
space. Since gravitational force is independent of velocity,
no particle will experience it in the conformal   space. 
In other words, gravitational force is effectively removed
to some extent by dilation given in equation (\ref{CON}).
Consequently, if a particle is moving in a circular 
trajectory, then the only force acting on it is the 
centrifugal force.

If $\xi^a$ is a Killing vector in the original space then 
 $\xi^a$ is also a Killing vector in    the conformal space
if 
\begin{equation}
{\cal L}_{\xi}\ \hat{\phi} = 0
\end{equation}
This is trivially true in  coordinates adapted to Killing vector $ \xi^a$. Then the Killing
vectors in the original spacetime are also Killing vectors
in the conformal spacetime. Therefore, $ \hat{\xi^a} \ = \ 
(1,0,0,0) $ is the timelike Killing vector and  
 $ \hat{\eta^a} \ = \ (0,0,0,1) $ is the spacelike Killing 
vector which generates circular orbits in the conformal 
spacetime. The quasi-Killing trajectories
\begin{equation}
\hat{\chi}^a \ = \  \hat{\xi^a}  \ + \ \omega \ \hat{\eta^a}
\end{equation}
generate circular orbits and the only force acting on these 
particles are centrifugal force. It is easy to prove that 
the expression for centrifugal       force is now
\begin{equation}
\hat{Z}_a \ =  \  \hat{u}^b \hat{\nabla}_b \hat{u}_a 
\label{C:CFF}
\end{equation}
where 
\begin{equation}
 \hat{u}_a \ = \ e^{\hat{\psi}} \hat{\chi}_a \ \ {\textstyle
and } \ \ e^{-2\hat{\psi}} \ = \ \hat{\chi}_a \hat{\chi}^a
\end{equation}
%------------------------------------------------------------
\subsection{Gyroscopic Precession in The Conformal Space   }
%------------------------------------------------------------
The gyroscopic precession in the conformal   spacetime can be 
computed     exactly as before. The Frenet-Serret parameters for
circular quasi-Killing trajectories can be written as
\begin{eqnarray}
\hat{\kappa}^{2}   &  =   & -\frac{1}{4} \left(\frac{  \hat{g}^{11} \hat{{\cal A}} 
^{2}_{1} +
\hat{g}^{22} \hat{{\cal A}} ^{2}_{2} }{ \hat{{\cal A}} ^{2}} \right)  \label{C:K} \\
\hat{\tau}^{2}_{1}  &  =  &  \left(\frac{ \hat{{\cal B}} ^{2}}{4 \hat{\Delta}_{3} \left( \hat{g}^{11}\hat{{\cal A}} ^{2}_{1} + \hat{g}^{22}\hat{{\cal A}} ^{2}_{2}\right)}\right) \cdot 
\nonumber \\ 
\             &   \ & \cdot \left( \frac{\hat{g}^{11}\hat{{\cal A }}_{1}\hat{{\cal B}} _{1} + 
\hat{g}^{22}\hat{{\cal A }}_{2}\hat{{\cal B}} _{2}}{ \hat{{\cal B}} } - \frac{\hat{g}^{11}
\hat{{\cal A}}^ {2}_{1} + \hat{g^{22}} \hat{{\cal A}} ^{2}_{2}}{ \hat{{\cal A} }} 
\right)^{2} \label{C:T1} \\ 
\hat{\tau}^{2}_{2}  &  =  &  \frac{ \hat{g}^{11} \hat{g}^{22}  \left( \hat{{\cal A}}_{1} \hat{{\cal B}}
_{2} - \hat{{\cal A}}_{2} \hat{{\cal B}}_{1} \right)^{2}}{ 
4 \hat{\Delta}_{3} \left( \hat{g}^{11} \hat{{\cal A}}
^{2}_{1} + \hat{g}^{22} \hat{{\cal A}}^{2}_{2}
\right) }
 \label{C:T2}
\end{eqnarray}
where
\begin{eqnarray}
 \hat{{\cal A}} \ =  \ 
\hat{\xi}^a  \hat{\xi}_a \ + \ 
\omega^2 \hat{ \eta }^a \hat{ \eta }_a \ = \ \frac{{\cal A}}
{ \hat{\phi}} \nonumber \\
 \hat{{\cal B}} \ =  \ \omega \ \hat{ \eta }^a \hat{ \eta }_a
\ =  \ \frac{{\cal B}}{ \hat{\phi}}  \label{C:AB} \\
\hat{\Delta}_{3} \ = \ (\hat{\xi}^a  \hat{\xi}_a ) 
(\hat{ \eta }^b \hat{ \eta }_b)
\end{eqnarray}
and
\begin{eqnarray}
\hat{{\cal A}}_a   \ =  \ \omega^2 (\hat{ \eta }^b \hat{ \eta }_b )_{,a} \ \ \ ; a =1,2 \nonumber \\
\hat{{\cal B}}_a   \ =  \ \omega (\hat{ \eta }^b \hat{ \eta }_b )_{,a} \ \ \ ; a =1,2 
\end{eqnarray}
One can then show that
\begin{eqnarray}
\hat{{\cal A}}_a \ =  \ \frac{ \hat{\phi} {\cal A}_a -{\cal A}\hat{\phi}_{,a}}{ \hat{\phi}^2} \nonumber \\
\hat{{\cal B}}_a \ =  \ \frac{ \hat{\phi} {\cal B}_a -{\cal B}\hat{\phi}_{,a}}{ \hat{\phi}^2} \ =  \ \frac{ \hat{{\cal A}}_a}{\omega}
\end{eqnarray}
With the help of the  above equations, $\hat{\kappa}^2$ can be
related to $ {\kappa}^2$. After some simplification we have,
\begin{equation}
\hat{\kappa}^2 \ = \ \hat{\phi}{\kappa}^2  \ - \ \frac{1}
{4\hat{\phi}} ( g^{ab} \ \hat{\phi}_{,a} \ \hat{\phi}_{,b} )
\ + \ \frac{1}{2 {\cal A}} ( g^{ab} \ {\cal 
A}_{a} \ \hat{\phi}_{,b} )
\end{equation}
From the definition of $\hat{\kappa} $ and the expression for
the centrifugal       force as in (\ref{C:CFF}), it is clear that 
the two are one and the same. This is because the
 contribution from the gravitational force has been removed
and the acceleration that appears is due to the centrifugal
force alone. We can relate $ \hat{\tau}^2_1 $  to 
$ \hat{\kappa}^2 $ by using the expression for
$ \hat{\tau}^2_1 $  to obtain 
\begin{equation}
\hat{\tau_1}^2 \ = \ - \frac{\hat{\kappa}^2}{ \hat{\Delta}_3
\omega^2} \label{C:con1}
\end{equation} 
 The  above equation is similar to equation(\ref{conection}) 
which relates $\tau_1$ to the centrifugal    force. It can also
be shown that
\begin{equation}
\hat{\tau}_2^2 \  = \ 0 \label{C:con2} 
\end{equation}
everywhere in the  conformal spacetime.

 From equations (\ref{C:con1}) and (\ref{C:con2})
it is clear that gyroscopic precession also reverses when
$ {\hat{\kappa}} \ = \ 0 $ and that in turn 
corresponds to the 
centrifugal force reversal. Also $ {\hat{\kappa}} \ = \ 0 $ 
corresponds to the geodesic condition in the conformal space, which 
represents the null geodesics in the original space as 
given in reference \cite{ACL}. 

To sum up, we have factored out the contribution due to the
gravitational   force   by conformal transformation and have 
shown in a simple manner the simultaneous reversal of both
the gyroscopic precession and the centrifugal       force
at the photon orbit.
%------------------------------------------------------------
%------------------------------------------------------------
\section{ Gravi-electric   and  Gravi-magnetic   fields}
%------------------------------------------------------------ %------------------------------------------------------------
Gravielectric and gravi-magnetic       fields are closely
related to the idea of inertial forces. These fields 
with respect to observers following the integral curves of 
$n^a$  can be defined as follows. \newline 
Gravi-electric    field:
\begin{eqnarray}
  E^a \ = \ F^{ab} n_b 
\end{eqnarray} 
 Gravi-magnetic   field:
\begin{eqnarray}
  H^a \ = \ \tilde{F}^{ab} n_b 
\end{eqnarray}
where $\tilde{F}^{ab} $ is the dual of  $F^{ab}$,
\begin{equation}
\tilde{F}^{ab} \ = \  \frac{1}{2} (\sqrt{-g})^{-1} 
\varepsilon ^{abcd} F_{cd} 
\end{equation}
In the above, as before, $ F^{ab}= e^\psi(\xi_{a;b} + \omega
\eta_{a;b} ) $.
The equation of motion   is 
\begin{equation}
\dot{u}^a \ = \ F^{ab} u_b
\end{equation}
Projecting onto the space orthogonal to $n^a$ with $ h_{ab}
= g_{ab} - n_a n_b $ and decomposing $ u_a $ as given in
(\ref{DEC}), we get 
\begin{equation}
\dot{u}_{\bot a} \ = \ \gamma \left[ F_{ac}n^c \ + \ v 
(F_{ac}\tau^c \ - \ n_a F_{bc} n^b \tau^c ) \right] 
\end{equation}
where $\gamma $ is the normalization factor. This equation can 
be written in the form 
\begin{equation}
\dot{u}_{\bot a} \ = \ \gamma \left[ F_{ac}n^c \ + \ v
\sqrt{-g} \varepsilon_{abcd} n^b \tau^c H^d \right],
\end{equation} 
or
\begin{equation}
\dot{u}_{\bot a} \ = \ \gamma \left[ E \ + \ v \times H 
\right]
\end{equation}
We can therefore define \newline
Gravi-electric   force:
\begin{equation}
 f_{GEa} \ = \ \gamma F_{ac}n^c 
\end{equation}
Gravi-magnetic   force:
\begin{equation}
f_{GHa} \ = \ \gamma  v \sqrt{-g} \varepsilon_{abcd} n^b \tau^c H^d  \ = \
\gamma v (F_{bc}\tau^c \ - \ n_a F_{bc} n^b \tau^c ) 
\end{equation}
%------------------------------------------------------------
\subsection{Relations among  Gravi-electric  , Gravi-magnetic  
and Inertial forces}
%------------------------------------------------------------
\subsubsection{Static Case}
%------------------------------------------------------------
We have defined the gravi-electric field $E_a$ by
\[
\gamma E_a = \gamma F_{ac} n^c 
\]
If we substitute for $  F_{ab }= e^\psi(\xi_{a;b} + \omega
\eta_{a;b} ) $, we get
\begin{equation}
f_{GEa} = \gamma E_a = \gamma F_{ac} n^c = - e ^{2( \psi+\phi)} G_a  
\end{equation}
So,
\begin{equation}
E_a = - e ^{( \psi+\phi)} G_a 
\end{equation}
Here $G_a $ is the gravitational force.
Similarly we have for the gravi-magnetic       field
\[
f_{GHa} =  \gamma v ( F_{ac} \tau^c - n_a n^b F_{bc} \tau^c )
\]
The second term  in this equation is identically zero because
the Killing vector field $\xi^ a $ and $ \eta^a $ commute
and we get
\begin{eqnarray}
f_{GHa} & = & \gamma  v \sqrt{-g} \varepsilon_{abcd} n^b \tau^c H^d  \nonumber \\
\ & = & \  \gamma  v  F_{ac} \tau^c \nonumber \\
\ & = & \  \left[ e^{2(\psi + \alpha )} \omega^2 G_a - Z_a 
\right]
\end{eqnarray}  
The above relation clearly shows the connection between the
gravi-magnetic       force on the one hand and the 
gravitational and centrifugal forces on the other.
%-------------------------------------------------------------
\subsubsection{Stationary Case}
%-------------------------------------------------------------
In the stationary case, $n^a$ is given by equation (\ref{NA}). As before we decompose $ \omega = \tilde{\omega} + \omega_0 $,
where $\omega_0$ is given by (\ref{OM}). Then a 
straightforward computation gives the expression for the gravi-electric field.
\begin{equation}
E_a = -e^{(\psi+\phi)} G_a \ + \ e^{-(\psi+\phi)} C_a 
\end{equation}
and the gravi-electric force,
\begin{equation}
f_{GEa} = \gamma E_a = -e^{2(\psi+\phi)} G_a \ + \ C_a 
\end{equation}
This shows the relation of gravi-electric field or force to
both  gravitational and centrifugal forces. In the stationary
case also we have
\begin{equation}
n_a n^b F_{bc} \tau^c \ \equiv \  0
\end{equation}
Then   it follows
\begin{eqnarray}
f_{GHa} & \equiv & \gamma v \sqrt{-g} \varepsilon_{abcd}
n^d \tau^c H^d \nonumber \\
\ & = & \  \gamma v  F_{ac} \tau^c \nonumber \\
\ & = & \  \left[\frac{C_a}{2} \ + \ 
e^{2(\psi+\alpha)} \tilde{\omega}^2  G_a \ - \ Z_a  \right]
\end{eqnarray} 
Hence gravi-magnetic force is related to all the three 
inertial forces -- gravitational, centrifugal and Coriolis.
%------------------------------------------------------------
\subsection{Gravi-electric and Gravi-magnetic fields with
respect to comoving frame}
%----------------------------------------------------------
In the previous section we have defined gravi-electric and
gravimagnetic fields with respect to the 
irrotational congruence.
Similarly these fields can be defined with respect to the 
four velocity $u^a $ of the particle as follows. \newline
Gravi-electric field:
\begin{equation}
\tilde{E^a} \ = \  F^{ab} u_b
\end{equation}
Gravi-magnetic field:
\begin{equation}
\tilde{H}^a \ = \ \tilde{ F}^{ab} u_b
\end{equation}
Where $\tilde{ F}^{ab}$ is dual to $F^{ab}$ as before. The 
equation of  motion takes the form
\begin{equation}
a^a \ = \ \tilde{E}^a 
\end{equation}
Precession frequency can be written simply as
\begin{equation}
\omega^a \ = \ \tilde{H}^a
\end{equation}
Following Honig, Sch\"{u}cking and Vishveshwara \cite{HSV},
Frenet-Serret parameters $\kappa, \tau_1 $ and $\tau_2 $
can be expressed in terms of gravi-electric and 
gravi-magnetic fields.
\begin{equation}
\kappa \ = \ | \tilde{E} | 
\end{equation}
where
\begin{equation}
| \tilde{E} | \ = \ \sqrt{ -\tilde{E}^a\tilde{E}_a }
\end{equation}
\begin{equation}
\tau_1 \ = \  \frac{ |\tilde{P} | }{| \tilde{E} |}
\end{equation}
where
\begin{eqnarray}
\tilde{P}^a \ = \ \varepsilon^{abcd}\tilde{E}_b\tilde{H}_a
u_d \ = \ \tilde{E}\ \times \tilde{H} \\
| \tilde{P} | \ = \ \sqrt{ -\tilde{P}^a\tilde{P}_a }    
\end{eqnarray}
and
\begin{equation}
\tau_2 \ = \ - \frac{\tilde{H}^a \tilde{E}_a }
{| \tilde{E} |}
\end{equation}
Frenet-Serret tetrad components can also be expressed 
in terms of $\tilde{E}^a, \  \tilde{H}^a $ and 
$\tilde{P}^a$,  
\begin{eqnarray}
e^a_{(1)}  & = & \frac{\tilde{E}^a}{ |\tilde{E} |} 
\nonumber \\
e^a_{(2)}  & = & \frac{\tilde{P}^a}{ |\tilde{P} |} \\
e^a_{(3)}  & = & \frac{ \varepsilon^{abcd}\tilde{E}_b 
\tilde{P}_c u_d }{ \tilde{P}^r \tilde{E}_r }
\nonumber
\end{eqnarray}
In reference \cite{HSV}, these expressions had been derived
for charged particle motion in a constant electromagnetic 
field. We have now demonstrated  
the exact analogues in the case of gravi-electric 
and gravi-magnetic fields. The one-to-one correspondence is
indeed remarkable.

Several authors have discussed gravi-electromagnetism in
earlier papers with application to gyroscopic precession.
We may cite as examples the papers by Embacher\cite{FE}, 
Thorne and Price\cite{TP}, Jantzen, Carini and Bini\cite{JCB},
and Ciufolini and Wheeler\cite{CW}.
%----------------------------------------------------------
%----------------------------------------------------------
\section{Conclusion}
%----------------------------------------------------------
%----------------------------------------------------------
The main purpose of the present paper was to establish
covariant connection between gyroscopic precession on the 
one hand and the analogues of inertial forces on the other.
This  has been accomplished in the case of axially
symmetric stationary spacetimes  for circular orbits.
In the special case of static spacetimes gyroscopic 
precession can be  directly related to the centrifugal 
force. From this we have been able to prove that both  
precession and  centrifugal force reverse at a photon
orbit, provided the latter exists. In the case of 
stationary spacetimes, the corresponding relations are more
complicated. The  place of centrifugal force is now taken 
by a combination of centrifugal and Coriolis-Lense-Thirring
forces. As a result, gyroscopic precession and centrifugal 
force do not reverse in general at  the photon orbit.
We have also studied some of the above aspects in the 
spacetime conformal to the original static spacetime. In 
this approach  part of the gravitational effect is
factored out thereby  achieving certain degree of 
simplicity and transparency in displaying interrelations
and the reversal phenomenon. Closely related to these considerations is the idea of gravi-electric and 
gravi-magnetic  fields. We have covariantly  defined these
with respect to the  globally  hypersurface orthogonal 
vector field that constitutes the general relativistic 
equivalent of Newtonian rest frame.  In this instance, 
these fields can be related to the inertial forces. When
these fields are formulated with respect to the orbit
under consideration, they lead to a striking similarity
to the corresponding  physical 
quantities that arise for a charge  moving in an actual, 
constant electromagnetic field. We have thus established 
connections and correspondences among several interesting 
general relativistic phenomena.

%----------------------------------------------------------
%----------------------------------------------------------

\end{document}